\documentclass[aps,graphicx,twocolumn]{revtex4}%
\usepackage{url}
\usepackage{appendix}
\usepackage[colorlinks]{hyperref}
\hypersetup{%
    plainpages=true,
    breaklinks=true,
    hypertexnames=false,
    pageanchor=true,
    colorlinks=true,
    linkcolor={blue},
    citecolor={blue},
    urlcolor={blue},
    anchorcolor={black}
}

\usepackage{amsmath}
\usepackage{graphicx}
\usepackage{subfigure}

\def \GHZ{\text{GHZ}}
\def \eV{\,\text{eV}}

\newcommand{\ket}[1]{\mbox{$|#1\rangle$}}

\begin{document}


\title{Resource-efficient analyzer of Bell and Greenberger-Horne-Zeilinger
states of multiphoton systems}

\author{Tao Li,$^{1,2}$ Adam Miranowicz,$^{2,3}$ Keyu Xia,$^{4}$ and Franco Nori$^{2,5}$}
\affiliation{$^{1}$School of Science, Nanjing University of Science and Technology, Nanjing 210094, China\\
$^{2}$Theoretical Quantum Physics Laboratory, RIKEN Cluster for Pioneering Research, Wako-shi, Saitama 351-0198, Japan\\
$^{3}$Faculty of Physics, Adam Mickiewicz University, 61-614 Pozna\'{n}, Poland\\
$^{4}$College of Engineering and Applied Sciences, Nanjing University, Nanjing 210093, China\\
$^{5}$Physics Department, The University of Michigan, Ann Arbor,
Michigan 48109-1040, USA}

\date{\today }

\begin{abstract}
We propose a resource-efficient error-rejecting entangled-state
analyzer for polarization-encoded multiphoton systems. {Our
analyzer is based on two single-photon quantum-nondemolition
detectors, where each of them is implemented with a four-level
emitter (e.g., a quantum dot) coupled to a one-dimensional system
(such as a micropillar cavity or a photonic nanocrystal
waveguide).} The analyzer works in a passive way and can
completely distinguish $2^n$ Greenberger-Horne-Zeilinger~(GHZ)
states of $n$ photons without using any active operation or fast
switching. The efficiency and fidelity of the GHZ-state analysis
can, in principle, be close to unity, when an ideal single-photon
scattering condition is fulfilled. For a nonideal scattering,
which typically reduces the fidelity of a GHZ-state analysis, we
introduce a passively error-rejecting circuit to enable a
near-perfect fidelity at the expense of a slight decrease of its
efficiency. Furthermore, the protocol can be directly used to
perform a two-photon Bell-state analysis. This passive,
resource-efficient, and error-rejecting protocol can, therefore,
be useful for practical quantum networks.
\end{abstract}

\maketitle

\section{Introduction}

Quantum entanglement is a fascinating phenomenon in quantum
physics~\cite{Horodecki2009review}, which provides a promising
platform for various quantum technologies, including quantum
communication networks~\cite{kimble2008quantum}.
 {Sharing quantum entanglement among distant network nodes is
a prerequisite for many practical
applications}~\cite{Cirac99Distributed, Lim2005Repeat,
Jiang2007Distributed,
Qin2017Heralded,lo2014secure,bennett1993teleporting,pirandola2015advances}.
There are two main obstacles for practical applications of
multipartite quantum entanglement, i.e., entanglement generation
{over} desired nodes and entanglement analysis within a local
node.  {Usually, it is difficult to distribute local
entanglement over spatially-separated nodes due to channel high
losses~\cite{Gisin2002RMP}. An efficient method to overcome
channel noise uses quantum repeaters~\cite{dur1999quantum,
jiang2009quantum, Wang2012QR, munro2012quantum}, which are based
on entanglement purification~\cite{Bennett96EPP, Deutsch96EPP,
Sheng2008EPP, Sheng2010DEPP} and quantum
swapping~\cite{Swap1993PRL, chen2011hybrid,
Hu2011PRB,Chen2013Swaping, Su2016swap}.} By applying a proper
entanglement analysis and local operations, one can complete an
{entanglement-purification protocol} to distill some entanglement
of a higher fidelity, and
 enlarge the distance of an entangled channel through quantum
swapping. {In addition to entanglement purification and
entanglement swapping,
 Bell-state analysis is crucial, e.g., for quantum
teleportation~\cite{bennett1993teleporting,pirandola2015advances},
quantum secure direct communication~\cite{long2002theoretically,
Deng2003Two-step, hu2016experimental, zhang2017quantum}, and
quantum dense coding~\cite{pan2012multiphoton}.} It plays an
essential role in various entanglement-based quantum information
processing protocols~\cite{lo2014secure,bennett1993teleporting,
pirandola2015advances, long2002theoretically, Deng2003Two-step,
hu2016experimental, zhang2017quantum, Gisin2002RMP,
dur1999quantum, jiang2009quantum, Wang2012QR, munro2012quantum,
Bennett96EPP, Deutsch96EPP, Sheng2008EPP, Sheng2010DEPP,
Swap1993PRL, chen2011hybrid, Hu2011PRB,Chen2013Swaping,
Su2016swap,pan2012multiphoton}.

{Multipartite entanglement, compared to two-particle entanglement,
is more powerful to reveal the nonlocality of quantum
physics~\cite{Horodecki2009review,pan2012multiphoton,
Deng201746,Tashima2016multipartie}. The
Greenberger-Horne-Zeilinger (GHZ) states enable more refined
demonstrations of quantum nonlocality, and  can be used to build
more complex quantum networks involving many
nodes~\cite{Hillery99QSS, Schauer2010QSS-Exp, farouk2018robust,
dong2017fault} and to perform, i.e., conference-key
agreement~\cite{Ribeiro2018conference}. Furthermore, GHZ states
enable efficient methods for large-scale cluster-state generation
for measurement-based quantum
computing~\cite{briegel2009measurement,Tanamoto2006Cluster,You2007Cluster,
tanamoto2009efficient, Economou2010Cluster, Gimeno2015GHZ-QC,
Liying2015GHZ-QC}, and  also provide a useful basis for quantum
metrology~\cite{giovannetti2004quantum,Dur2014MetrologyQEC}.}
The generation and analysis of $n$-photon GHZ entanglement are
highly demanding. To date, various efficient methods to generate
the GHZ entanglement have been developed for different physical
systems~\cite{wei2006generation, wang2010one, Anton2018GHZ,
monz2011-14qubit, Kaufmann2017ion-GHZ, Yu2007GHZ,
zheng2012generation, Reiter2016GHZ, shao2017dissipative}. {In
photonic systems, an eight-photon GHZ state and a three-photon
high-dimensional  GHZ state have been experimentally
demonstrated~\cite{huang2011experimental8,
yao2012observation8,malik2016multi,Krenn2016Automated} by
performing quantum fusion combined with post-selection operations
and quantum interference}~\cite{pan2012multiphoton,
Krenn2017PathGHZ, Bergamasco2017pathGHZ}. By using a time delay, a
resource-efficient method was proposed and demonstrated~\cite{Megidish2012Resource} for generating a
six-photon GHZ state.
It is possible to generate photonic GHZ states or other
multipartite-entangled states in a deterministic way {based
on nonlinear processes}~\cite{duan2004scalable, li2011photonic,
hao2015quantumRy, reiserer2014quantum, reiserer2015cavity}.
However, it is {difficult} to distribute such a GHZ state
efficiently to distant nodes, due to the inefficiency of the GHZ
sources and high { losses during transmission}~\cite{Gisin2002RMP,
perseguers2013distribution}. One possible solution is to establish
entanglement pairs between a center node and distant nodes in
parallel~\cite{dur1999quantum, jiang2009quantum, Wang2012QR,
munro2012quantum}, and then to perform  quantum swapping
with a GHZ-state analysis in the center
node~\cite{bose1998multiparticle, Lu2009swap}.

In 1998, Pan and Zeilinger proposed, {to our knowledge,} the first
practical GHZ-state analysis with linear-optical
elements~\cite{Pan1998GHZanalyzer}. {Their proposal} can identify
two of \emph{n}-photon GHZ states by post-selection operations. In
principle, one can constitute a nearly deterministic
\emph{n}-photon GHZ-state analysis with linear optics, when
massive ancillary photons are used~\cite{Kop2007linear}. However,
according to the Cansamiglia-L\"utkenhaus no-go
theorem~\cite{Calsamiglia2001}, perfect and deterministic
Bell-state analysis on two polarization-encoded qubits is
impossible by using only linear-optical elements (in addition to
photodetectors) and auxiliary modes in the vacuum state.
By taking into account nonlinear processes, a complete GHZ-state analysis for photonic
systems becomes possible~\cite{qian2005universal,
xia2014complete}, and {can achieve perfect efficiency and
fidelity} for an ideal process. {Moreover, a complete entangled-state
analysis for hyperentangled or redundantly
encoded photon pairs  is possible}~\cite{Zhou2015logic, sheng2010BEll,
ren2012complete, Wang2012analysis, Liu2015GenerationNV, Wang:16}.
The existing GHZ-state analyses typically require active
operations and/or fast switching, and  always require more
quantum resources when the photon number of { a given GHZ state
increases}. Furthermore, the fidelity of the Bell-state or
GHZ-state {analyses significantly depends on the nonlinearity
strength of realistic} nonlinear
processes~\cite{reiserer2015cavity}. A deviation from an ideal
nonlinear process leads to errors and, thus, reduces the
fidelity. These disadvantages significantly limit
 applications of a GHZ-state analysis for practical
quantum networks.

Here we propose a resource-efficient passive protocol of a
 multiphoton GHZ-state analysis using only {two
single-photon nondestructive [quantum nondemolition (QND)]
detectors, three standard (destructive) single-photon detectors},
and some linear-optical elements. The GHZ-state analysis circuit
is universal, and can completely distinguish {$2^n$ GHZ states with
different photon numbers $n$,} according to the measurement results of
{single-photon nondestructive and destructive} detectors. The circuit works in a passive
way as the Pan-Zeilinger GHZ-state analyzer
does~\cite{Pan1998GHZanalyzer}. During
the entangled-state analysis, there are neither active operations
on ancillary atoms nor adaptive switching of
photons~\cite{witthaut2012photon}. The efficiency of our GHZ-state
analysis {can, in principle, be equal to one.}
Moreover, {our protocol has no requisite for direct Hong-Ou-Mandel
interference which requires simultaneous operations} on two
individual photons. Thus, we can significantly simplify the
process of   GHZ-state analysis  and, subsequently, the structure
of {multinode} quantum networks. Furthermore, the detrimental
effect on the fidelity, introduced by a nonideal scattering
process, can be eliminated passively at the expense of a decrease
of its efficiency. {Therefore, our protocol is resource efficient
and passive, and can be used to efficiently entangle distant nodes
in complex quantum networks.}

The paper is organized as follows: {A quantum interface between a
single photon and a single quantum dot~(QD) is introduced briefly
in Sec.~\ref{sec.II} for performing QND  measurements on}
linearly polarized photons.  {In Sec.~\ref{sec.III}, a passive
GHZ-state analysis circuit is presented. In Sec.~\ref{sec.IV}, a
method to efficiently generate entanglement among distant nodes is
described. Subsequently, the performance of the circuit, with
state-of-the-art experimental parameters, is discussed in
Sec.~\ref{sec.V}. We conclude with brief discussion and conclusions
in Secs.~\ref{sec.VI} and  \ref{sec.VII}. Moreover,
Appendixes~\ref{app-Bell} and~\ref{app-ghz3} present the two simplest examples of our method
for the analysis of two-photon Bell states and three-photon GHZ
states.}

\begin{figure}[!tpb]
\begin{center}
\includegraphics[width=7.2 cm, angle=0]{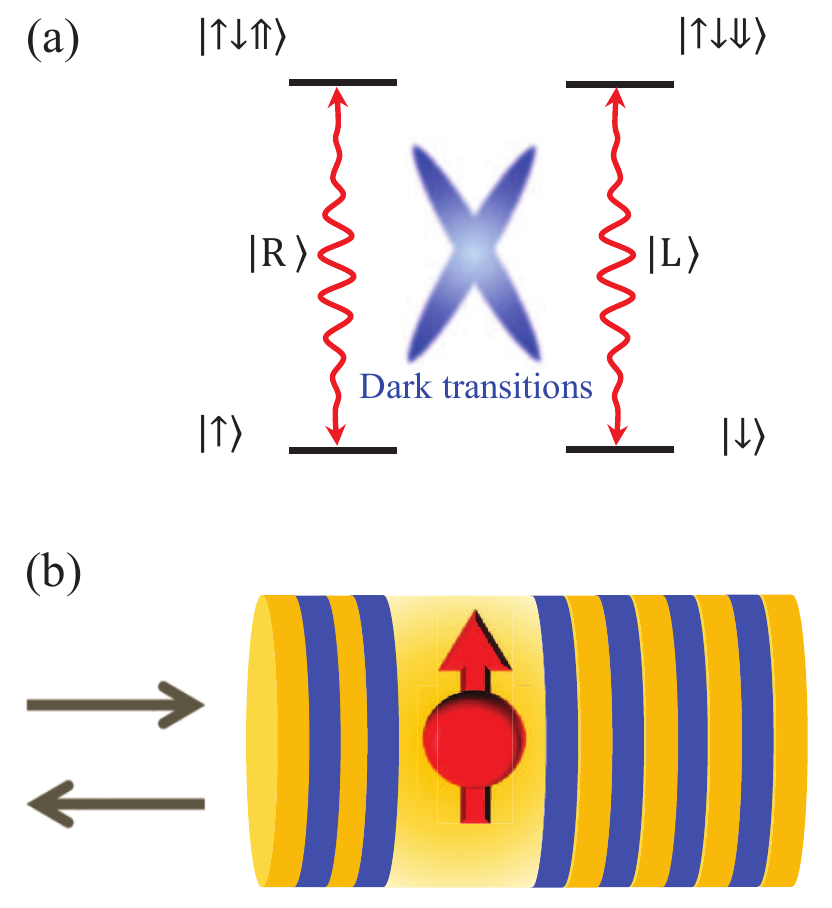}
\caption{{Proposal of quantum-nondemolition detection
based on} spin-dependent transitions for the negatively charged
exciton $X^-$. (a) Relative-level structure and optical transition
of a singly charged quantum dot~(QD); (b) a QD coupled to an
optical micropillar cavity.
 {Here, $|\!\uparrow\rangle$ ($|\!\downarrow\rangle$) denotes
the electron spin state with $J_z$ equal to 1/2   ($-$1/2), and
$|\!\!\uparrow\downarrow\Uparrow\rangle$
($|\!\!\uparrow\downarrow\Downarrow\rangle$) denotes
the trion state of $X^-$ with $J_z$ equal to 3/2   ($-$3/2). A
photon in a right- (left-) circularly-polarized state $|R\rangle$
($|L\rangle$}) can only couple to the transition
$|\!\uparrow\rangle\leftrightarrow|\!\uparrow\downarrow\Uparrow\rangle$
($|\!\downarrow\rangle\leftrightarrow|\!\uparrow\downarrow\Downarrow\rangle$).
Therefore, the cross transitions are forbidden by the
quantum-optical selection rules.
 }
\label{fig1}
\end{center}
\end{figure}

\section{Single-photon QND detector}
\label{sec.II}

An efficient {interface, between a single photon and a single
emitter,} constitutes a necessary building block for various kinds
of quantum tasks, especially for long-distance or distributed
quantum networks~\cite{kimble2008quantum, reiserer2015cavity}. To
begin with, we consider a process of single-photon scattering by a
four-level emitter coupled to a one-dimensional system, such as a
QD coupled to a micropillar cavity or a {photonic nanocrystal}
waveguide~\cite{reithmaier2004strong, Arcari14Near-Unity,
li2018gate, lodahl2015interfacing, hu2008giant,
hu2008deterministic}. A singly-charged self-assembled In(Ga)As QD
has four energy levels~\cite{hu2008giant, hu2008deterministic,
warburton2013single}: two ground states of $J_z=\pm1/2$, denoted
as $|\!\uparrow\rangle$ and $|\!\downarrow\rangle$, respectively;
and two optically excited trion states $X^-$, consisting of two
electrons and one hole, with $J_z=\pm3/2$, denoted as
$|\!\!\uparrow\downarrow\Uparrow\rangle$ and
$|\!\!\uparrow\downarrow\Downarrow\rangle$, respectively. Here the
quantization axis $z$ is along the growth direction of the QD and
it is the same as the direction of the input photon. Therefore,
there are two circularly-polarized dipole transmissions which are
degenerated when the environment magnetic field is zero, as shown
in Fig.~\ref{fig1}. A right-circularly polarized photon
$|R\rangle$ and a left-circularly polarized photon $|L\rangle$ can
only couple to the transitions
$|\!\uparrow\rangle\leftrightarrow|\!\uparrow\downarrow\Uparrow\rangle$
and
$|\!\downarrow\rangle\leftrightarrow|\!\uparrow\downarrow\Downarrow\rangle$,
respectively.

The single-photon scattering process of a QD-cavity unit is
dependent on the state of the QD. There are two individual cases:
(1) If an input photon does not match the circularly polarized
transition of the QD, the photon excites the cavity mode that is
orthogonal to the polarization transition of the QD, and it is
reflected by a practically empty cavity with a loss probability
caused by photon absorption and/or side leakage. {(Hereafter, for
brevity,   we refer to the side leakage only, but we also mean
other photon absorption losses.)} However, (2) if an input photon
matches a given transition of the QD, the photon interacts with
the QD and
 is reflected by the cavity that couples to the QD.
{Therefore, a} \emph{j}-circularly polarized photon (where
$j=$right or left) in the input mode $\hat{a}^{\dagger}_{\omega_j,
\text{in}}$ of frequency $\omega_j$, after it is scattered by a
QD-cavity unit, evolves into an output mode
$\hat{a}^{\dagger}_{\omega_j, \text{out}}$ as
follows~\cite{hu2008giant, hu2008deterministic,
lodahl2015interfacing, warburton2013single}:
\begin{eqnarray} 
\hat{a}^{\dagger}_{\omega_j, \text{in}}|0, 0,
{\bar{s}}\rangle\;\rightarrow\; r_0\hat{a}^{\dagger}_{\omega_j, \text{out}}|0, 0, {\bar{s}}\rangle, \nonumber\\
\hat{a}^{\dagger}_{\omega_j, \text{in}}|0, 0, s\rangle
\;\rightarrow\; r_1\hat{a}^{\dagger}_{\omega_j, \text{out}}|0, 0,
s\rangle, \label{stateTransform}
\end{eqnarray}
where the state $|0, 0, s\rangle$~($|0, 0, \bar{s}\rangle$)
denotes   that both input and output fields are in the vacuum
state and the QD is in the state $|s\rangle$~($|\bar{s}\rangle$)
that {couples (does not couple)} to the input photon.
 {Under the assumptions of both adiabatic evolution of the cavity field and negligible excitation of the QD,}
the state-dependent reflection amplitudes $r_0$ and $r_1$,
corresponding, respectively, to the aforementioned cases (1) and
(2), are given by~\cite{hu2008giant, hu2008deterministic,
lodahl2015interfacing, warburton2013single}:
\begin{eqnarray} 
 r_0(\omega)&=&1-\frac{{\kappa}}{i(\omega_{c}\!-\!\omega)\!+\!\frac{\kappa}{2}\!+\!\frac{\kappa_s}{2}},\nonumber\\
\nonumber\\
r_1(\omega)&=&1-\frac{\kappa f}
{\!\!\!\left[i(\omega_{c}\!-\!\omega)\!+\!\frac{\kappa}{2}\!+\!\frac{\kappa_s}{2}\right]\!f+\!g^2},
\label{rcoeS}
\end{eqnarray}
where the auxiliary function $f$ is given by
$f=i(\omega_{X^-}-\omega)+\frac{\gamma}{2}$.
Here {$\omega_{X^-}$ is the transmission frequency of the QD and
$\omega_{c}$ is the resonant frequency of the cavity. These
frequencies} can be tuned to be equal to
$\omega_{X^-}=\omega_{c}$,   for simplicity. {Moreover, $\kappa$
describes a directional coupling between the cavity modes and the
input and output modes; $g$ denotes the coupling between the QD
and cavity; $\kappa_s$ represents the cavity side-leakage rate,
and $\gamma$ is the trion decay rate.}  {These formulas for the reflection coefficients are valid in
general for both weak and strong couplings~\cite{OBrien2016}.}

 {For ideal scattering in the strong-coupling regime with $\kappa_s\ll\kappa$ and $\gamma,
\kappa\ll{}g$~(or in the high-cooperativity regime with $\kappa_s\ll\kappa$, $\gamma\ll{}g\ll\kappa$,
and $\gamma\kappa\ll{}g^2$)~\cite{OBrien2016},}
an input photon, that is resonant with a QD
transition, is deterministically reflected by the QD-cavity unit.
A $\pi$-phase (\emph{zero}-phase) shift is introduced to the
hybrid system consisting of a photon and the QD with
$r_0=-1$ {for $g=0$~($r_1=1$ for $\kappa\gamma\ll{}g^2$),} if the photon decouples~(couples) to a
transition of the QD. When the QD is initialized to be in the
superposition state
$|\pm\rangle=(|\!\uparrow\rangle\pm|\!\downarrow\rangle)/\sqrt{2}$,
an input photon in a linearly polarized state evolves as follows:
 \begin{eqnarray} 
|H\rangle|\pm\rangle\rightarrow|V\rangle|\mp\rangle, \nonumber\\
|V\rangle|\pm\rangle\rightarrow|H\rangle|\mp\rangle.
\label{transformMatrix}
\end{eqnarray}
 {Equation~(\ref{transformMatrix}) means that if a QD receives
a single photon, then it receives the Pauli $\sigma_x$ unitary. On
the one hand, if the QD does not receive any photon, then it does
not change its state. Thus, if we can identify whether the QD
receives the Pauli $\sigma_x$ unitary, then it works as a QND
measurement for photons~\cite{witthaut2010photon,
hu2008giant,OBrien2016, reiserer2013nondestructive}.  Furthermore,
when the QD receives a photon, then it flips the polarization
state of the photon simultaneously~\cite{Li2012Robust,
Lit2016Error}. We will show in Sec.~\ref{sec.V} that the QND
measurement can work faithfully with a limited efficiency for
practical scattering, i.e., when $r_1(\omega)$ and $r_0(\omega)$
significantly deviate from their ideal values $\pm1$.}

\begin{figure}[!tpb]
\begin{center}
\includegraphics[width=8.6 cm, angle=0]{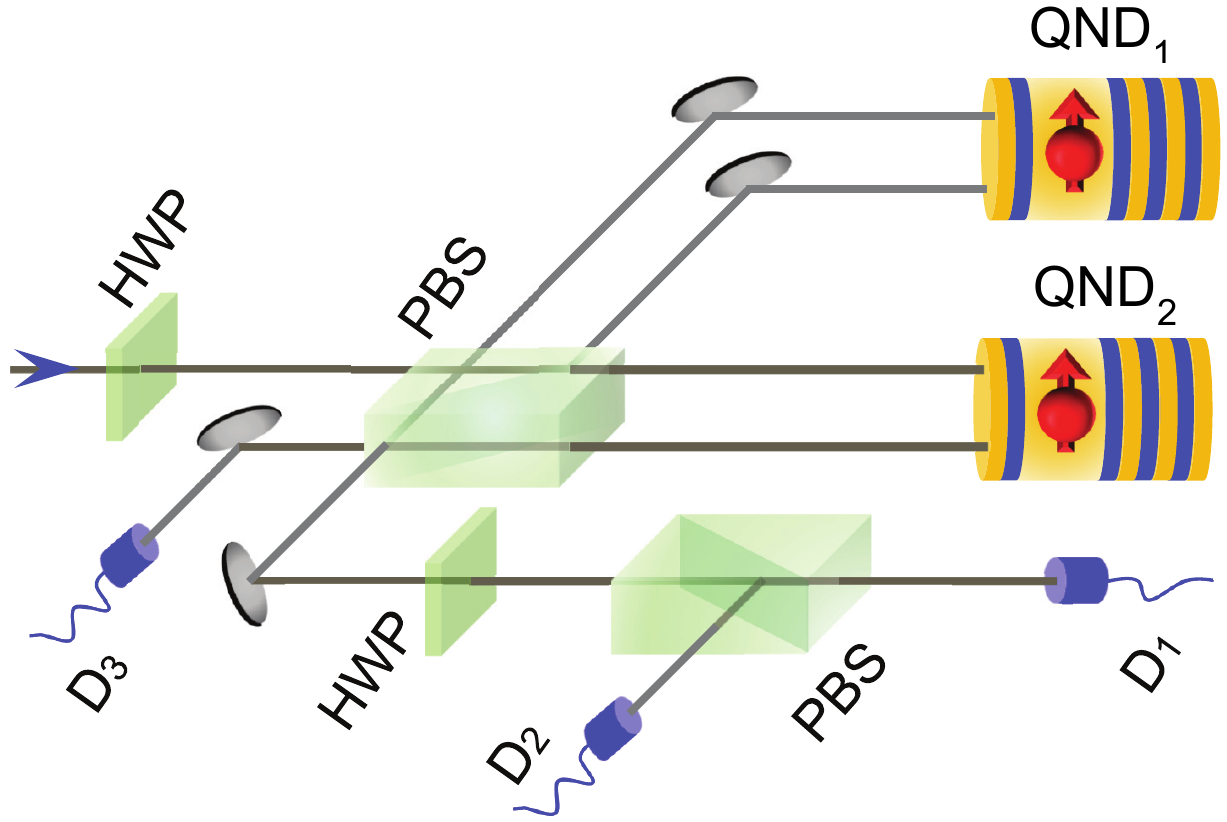}
\caption{ Schematics of the passive optical GHZ-state analyzer
using single-photon QND detectors. Here PBS denotes a {polarizing
beam splitter, which transmits photons with horizontal
polarization $|H\rangle$ and reflects those with vertical
polarization $|V\rangle$.} HWP represents a half-wave plate that
performs the Hadamard transformation on photons passing it, i.e.,
$|H\rangle\rightarrow(|\!H\rangle+|\!V\rangle)/\sqrt{2}$ or
$|V\rangle\rightarrow(|\!H\rangle-|\!V\rangle)/\sqrt{2}$. QND
detection completes a nondestructive measurement on single
photons, and D$_i $ ($i=1, 2, 3$) is an ordinary (destructive)
single-photon detector.} \label{fig2}
\end{center}
\end{figure}

\section{Passive GHZ-state analyzer} \label{sec.III}
\subsection{The setup}

So far, we have described a QND detection of linearly polarized
single photons. In this section, we describe how to
incorporate a QND detector into the setup for the passive optical
GHZ-state analysis, as shown in Fig.~\ref{fig2}. The setup is
composed of two half-wave plates~(HWPs), two polarizing beam
splitters (PBSs), two single-photon QND detectors in the state
$|+\rangle_1|+\rangle_2$, and several standard (destructive)
single-photon detectors. The HWP is tuned to perform
 the Hadamard transformation on photons passing it, i.e.,
$|H\rangle\rightarrow(|\!H\rangle+|\!V\rangle)/\sqrt{2}$ or
$|V\rangle\rightarrow(|\!H\rangle-|\!V\rangle)/\sqrt{2}$. The PBS
transmits linearly-polarized photons in the state $|H\rangle$ and
reflects photons in the state $|V\rangle$. The single-photon QND
and standard detectors complete the photon on-off measurements in
nondestructive and destructive ways, respectively.

\subsection{GHZ states}

For $n$-photon polarization-encoded GHZ states, the simplest two
can be expressed as~\cite{pan2012multiphoton}:
 \begin{eqnarray} 
|\GHZ_{00...0}\rangle&=&\frac{1}{\sqrt{2}}\left(|H\rangle^{\otimes{}n}+|V\rangle^{\otimes{}n}\right), \nonumber\\
|\GHZ_{00...1}\rangle&=&\frac{1}{\sqrt{2}}\left(|H\rangle^{\otimes{}n}-|V\rangle^{\otimes{}n}\right),
\label{GHZn1}
\end{eqnarray}
 {where the last ($n$th) bit  in the subscript of $|\GHZ_{00...n}\rangle$ refers to the phase ($\pm$).
If a photon is determined in the state
$|H\rangle$ or $|V\rangle$, the remaining $(n-1)$ photons
collapse into the same polarization.} To constitute a
complete basis for the $n$-photon system, one should take the
remaining $(2^{n}-2)$ orthogonal basis states into consideration,
 {
 \begin{eqnarray} 
 |\GHZ_{i_1i_2...i_n}\rangle&=&\bigotimes_{j=1}^{n-1}\sigma^{i_j}_{x_j}\otimes\sigma^{i_n}_{z_n} |\GHZ_{00...0}\rangle\nonumber\\
 &=&i\sigma^{i_n}_{y_n}\bigotimes_{j=1}^{n}\sigma^{i_j}_{x_j} |\GHZ_{00...0}\rangle,
\label{GHZnn}
\end{eqnarray}
which can be generated from $|\GHZ_{00...0}\rangle$ by
performing a single-photon rotation on each photon, and
$$\bigotimes_{j=1}^{n}\sigma^{i_j}_{x_j}=\sigma_{x_1}^{i_1}\otimes\sigma_{x_2}^{i_2}\otimes...\otimes\sigma^{i_n}_{x_n}.$$
Here, the superscripts $i_1$, $i_2$, ..., $i_n$ $\in\{0, 1\}$,
the Pauli operators
$\sigma_{x_j}=|H\rangle_j\langle V|+|V\rangle_j\langle H|$ perform
a polarization flip on the $j$th photon with $j=1, 2, ...,n$;
$\sigma_{y_n}=-i(|H\rangle_n\langle V|-|V\rangle_n\langle{}H|$);
$\sigma_{z_n}=|H\rangle_n\langle H|-|V\rangle_n\langle{}V|$
performs a phase flip on the $n$th photon; and the relative
phase between the two components of Eq.~(\ref{GHZnn}) is
determined by $i_n$; i.e., $i_n=0$~($i_n=1$) leads to a relative
phase of $0$~($\pi$).}

\subsection{State transformations for the GHZ-state analysis}\label{GHZ-state-analysis}

Now we focus on completely distinguishing the aforementioned $2^n$
GHZ states, which is of vital importance for multiuser quantum
networks~\cite{Pan1998GHZanalyzer, Lu2009swap,
bose1998multiparticle}. According to stabilizer
theory~\cite{Gottesman1996QECC,Toth2005entanglement,Schmid2008Discriminating,Greganti2015Practical},
the $n$-photon  state  {$|\GHZ_{00...0}\rangle$,} given in
Eq.~(\ref{GHZn1}), is a stabilizer state that can be uniquely
defined by $n$ stabilizing operators $S_k$,
 \begin{eqnarray} 
S_k=\left\{
        \begin{array}{ll}
          \sigma_{x_1}\otimes\sigma_{x_2}\otimes...\otimes\sigma_{x_n}, & \hbox{$k=1$;} \\
          \sigma_{z_{k-1}}\otimes\sigma_{z_k}, & \hbox{$k=2, 3, ..., n.$}
        \end{array}
      \right.
      \label{stabilizers}
\end{eqnarray}
Here the operators $\sigma_{z_k}$
perform a phase flip on the $k$th photon
with $k=1, 2, ..., $ and $n$; there is an implicit identity
$I^{\otimes(n-2)}$ acting on the remaining photons that is
suppressed in $S_{k\geq2}$ for simplicity.

The  set of operators $S_1$, $S_2$, ..., $S_n$ forms a complete
set of commuting observables; the $2^n$ GHZ states  are common
eigenvectors of all $S_k$'s with different
eigenvalues~\cite{Toth2005entanglement}, i.e.,
$|\GHZ_{00...0}\rangle$ gives an eigenvalue $+1$ for all $S_k$'s.
Therefore,  we can measure the stabilizing operators $S_k$'s to
completely discriminate $2^n$ GHZ states of an $n$-photon system.

Here the $n$-photon observable $S_1$ corresponds to {the
measurement of}  the relative phase between the two terms in a GHZ
state and can be nondestructively measured by using two QND
detectors introduced in Sec.~\ref{sec.II};  $S_{k\geq2}$
corresponds to parity detection on the pair of $(k-1)$th and $k$th
photons and is measured with direct polarization measurements on
each photon scattered by the QND detectors.
 {To explain in detail our GHZ-state analysis, we use
the ket notation instead of the stabilizer codes, since the
stabilizer states change during the analysis.}

For clarity, we divide this GHZ-state analysis into several steps. Let us suppose that
there is a spatial separation between each two optical elements
such that all photons can pass a given optical element before
entering another element. Note this requirement is not necessary,
and we will demonstrate, in the next section, that our proposal
also works when each photon is passing one by one from the input
port to the output port  and is measured by a single-photon
destructive detector.

After passing $n$ photons though the HWP, the Hadamard
transformation is performed on each photon, and the $2^n$ GHZ
states are changed into superposition states of $2^{n-1}$ (out of
$2^n$ possible)  product states,  {each with an even~(odd)
number of $V$-polarized photons for
$|\GHZ_{i_1i_2...i_{n-1}0}\rangle$~($|\GHZ_{i_1i_2...i_{n-1}1}\rangle$)
.} For instance, the states
$|\GHZ_{i_1i_2...i_{n-1}0}\rangle$ and
$|\GHZ_{i_1i_2...i_{n-1}1}\rangle$, after the Hadamard
transformation of each photon, evolve into
 {\begin{eqnarray} 
 |\varPsi_1\rangle&=&\frac{1}{\sqrt{2^{n-1}}}\sum_{m=0}^{[\frac{n}{2}]}\sqrt{C^{2m}_n}|G^{i_1,...,i_{n-1}}_{2m}\rangle, \nonumber\\
|\varPhi_1\rangle&=&\frac{1}{\sqrt{2^{n-1}}}\sum_{m=1}^{[\frac{n+1}{2}]}\sqrt{C^{2m-1}_n}|G^{i_1,...,i_{n-1}}_{2m-1}\rangle,
\label{GHZn11}
\end{eqnarray} 
respectively.} Here $[x]$ is the integer value function that
rounds the number $x$ down to the nearest integer;
  {$C^m_n=\frac{n!}{m!(n-m)!}$ is the  binomial coefficient;}
 the state $|G^{i_1,...,i_{n-1}}_{m}\rangle$ is an $n$-photon superposition
state that contains $m$ $V$-polarized photons and $(n-m)$
$H$-polarized photons as follows:  
\begin{equation}
 |G^{i_1,...,i_{n-1}}_m\rangle=\frac{Z}{\sqrt{C^m_n}}\sum_{\substack{l_1,...,l_n\in\{0, 1\}}}  {\delta_{m,m'}} \bigotimes_{j=1}^{n}\sigma^{l_j}_{x_j} |H\rangle^{\otimes{}n},
\label{G-m2}
\end{equation}
where $m'=\sum_{j=1}^n l_j$ and $\delta_{m,m'}$ is the Kronecker delta.
The phase of each component is determined by the operator
$Z=\bigotimes_{j=1}^{n-1}\sigma^{i_j}_{z_{j}}$, which is
simplified to an identity operator when analyzing
$|\GHZ_{00...0}\rangle$ and $|\GHZ_{00...1}\rangle$.

\subsection{  Measurements for the GHZ-state analysis}

 {As follows from the above analysis, the relative phase of
$|\GHZ_{i_1i_2...i_n}\rangle$, which is determined by
$i_n$, can be read out by measuring the number of $V$-polarized
photons in the even-odd basis  {after applying the Hadamard transformation to $|\GHZ_{i_1i_2...i_n}\rangle$.}
This measurement can be completed
by a setup consisting of a PBS and two QD-cavity units (referred
to as QND detectors). As demonstrated in Sec.~\ref{sec.II}, a
linearly polarized photon, after being scattered by a QND
detector, changes its polarization state into an orthogonal state
and flips the state of the detector QD. After all photons are
either transmitted or reflected by the first PBS, and scattered by
the QND detectors,} the hybrid states of the two QDs and the $n$
photons, corresponding to the states
$|\GHZ_{i_1i_2...i_{n-1}0}\rangle$ and
$|\GHZ_{i_1i_2...i_{n-1}1}\rangle$, evolve, respectively, into
\begin{eqnarray} 
|\varPsi_2^e\rangle&=&\bigotimes_{j=1}^{n}\sigma_{x_j}|\varPsi_1\rangle|+\rangle_1|+\rangle_2, \nonumber\\
|\varPhi_2^e\rangle&=&\bigotimes_{j=1}^{n}\sigma_{x_j}|\varPhi_1\rangle|-\rangle_1|-\rangle_2,
\label{GHZn1e}
\end{eqnarray}
 {if $n$ is even, and into }
\begin{eqnarray} 
|\varPsi_2^o\rangle&=&\bigotimes_{j=1}^{n}\sigma_{x_j}|\varPsi_1\rangle|+\rangle_1|-\rangle_2, \nonumber\\
|\varPhi_2^o\rangle&=&\bigotimes_{j=1}^{n}\sigma_{x_j}|\varPhi_1\rangle|-\rangle_1|+\rangle_2,  
\label{GHZn1o}
\end{eqnarray}
if $n$ is odd.  {The combined states of the two QDs in QND detectors are different, and can be used to deterministically distinguish
$|\varPsi_1\rangle$ from $|\varPhi_1\rangle$ for both cases of even and odd $n$.}

To make this point clearer, we continue our analysis to
measure the parity of each photon pair $[k-1,k]$ for the case of
an arbitrary \emph{even} $n$. Now, photons in different
polarization states combine again at the first PBS, which is
followed by an HWP. The HWP completes the Hadamard transformation
on each photon passing through it and evolves the photonic
component of the hybrid states into its original GHZ state, up to
a  phase difference $\pi$. For the states $|\varPsi_2^e\rangle$ and
$|\varPhi_2^e\rangle$, given in Eq.~(\ref{GHZn1e}), they evolve
into
 {
 \begin{eqnarray} 
|\varPsi_3\rangle&=&\pm|\GHZ_{i_1i_2...i_{n-1}0}\rangle|+\rangle_1|+\rangle_2, \nonumber\\
|\varPhi_3\rangle&=&\pm|\GHZ_{i_1i_2...i_{n-1}1}\rangle|-\rangle_1|-\rangle_2.
\label{GHZn1e1}
\end{eqnarray}
Here $|\GHZ_{i_1i_2...i_{n-1}0}\rangle$ and
$|\GHZ_{i_1i_2...i_{n-1}1}\rangle$ are the $n$-photon GHZ states
given in Eq.~(\ref{GHZnn}); their sign is determined by the summation of
the first $(n-1)$ subscripts with $m''=\sum_{j=1}^{n-1} i_j$,
i.e., ``$+$" for even $m''$ and ``$-$" for odd $m''$.}
Subsequently, a photon-polarization measurement setup, consisting
of a PBS and two destructive single-photon detectors D$_{1}$ and
D$_{2}$,  is used to detect the polarization of each photon and
then divides the measurement results according to the number of
clicks of each detector, i.e., when $n$ $H$-polarized ($V$-polarized) photons are detected,
the $n$ {input photons are} {projected into} either
$|\GHZ_{00...0}\rangle$ or $|\GHZ_{00...1}\rangle$, which
can be  distinguished by detecting the state of the QD
in each QND detector.

It is seen that there is neither active feedback nor fast
switching operations involved in the
 entangled-state analysis. The setup works in a completely
passive  way, which is similar to that based on linear-optical
elements and single-photon detectors.
When $n=2$, the GHZ-state analysis
setup enables a passive Bell-state analysis for
 two-photon systems, which are typically denoted as
\begin{eqnarray} 
 |\phi^{\pm}\rangle&=&\frac{1}{\sqrt{2}}\left(|H\rangle|H\rangle\pm|V\rangle|V\rangle\right), \nonumber\\
 |\psi^{\pm}\rangle&=&\frac{1}{\sqrt{2}}\left(|H\rangle|V\rangle\pm|V\rangle|H\rangle\right).
\label{Bell}
\end{eqnarray}
Detailed analyses for $n=2$ and $3$ are presented in Appendixes~A
and B, respectively.

\section{Efficient distant multipartite entanglement generation for quantum networks} \label{sec.IV}

In quantum multinode networks, multipartite entanglement
 among many nodes is useful for practical quantum
communication or distributed quantum computation~\cite{Deng201746,
pan2012multiphoton}. A direct method for sharing the GHZ
entanglement among several distant nodes can be enabled by a
faithful entanglement distribution after locally generating the
GHZ entanglement. However, the efficiency of such a multipartite
entanglement distribution significantly decreases {with the
increasing photon number involved in the GHZ
entanglement}~\cite{Gisin2002RMP}. Furthermore, the {experimental
methods for generating  multiphoton GHZ entanglement are still
inefficient} due to the limited experimental technologies. A
significantly more efficient method for distant GHZ state
generation can be achieved by entanglement swapping. In the
following, we describe a scheme for the GHZ entanglement
generation among three stationary qubits, and these stationary
qubits can be atomic ensembles, nitrogen-vacancy~(NV) centers,
QDs, and other systems~\cite{buluta2011natural}.

\begin{figure}[!tpb]
\begin{center}
\includegraphics[width=8.6 cm, angle=0]{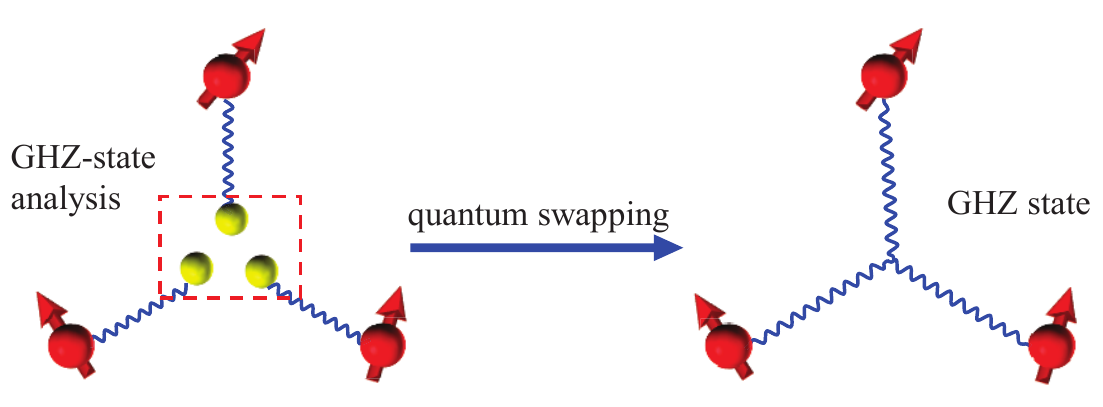}
\caption{ Schematics of nonlocal GHZ-state generation for
multiparty quantum networks. Here a red circle with an arrow
represents a stationary qubit, while a yellow circle represents a
photon. Each wave line represents entanglement between the
particles it connects. } \label{fig3}
\end{center}
\end{figure}

Suppose there are three communicating nodes in a quantum network,
say, Alice, Bob, and Charlie. An ancillary node~(Eve) shares
hybrid entanglement pairs with Alice, Bob, and Charlie,
respectively, as follows~\cite{, dur1999quantum, jiang2009quantum,
Wang2012QR, munro2012quantum, Bennett96EPP, Deutsch96EPP,
Sheng2008EPP, Sheng2010DEPP}:
 \begin{eqnarray} 
 |\phi\rangle_{ji}=\frac{1}{\sqrt{2}}\left(|\!\uparrow\rangle_j|H\rangle_{i}+|\!\downarrow\rangle_j|V\rangle_{i}\right),
\label{Bellh}
\end{eqnarray}
where the subscript $i$ (with $i=a, b, c$) represents the photons
owned by Eve, and it is entangled with the $j$th QD (with $j=A, B,
C$), which belongs to Alice, Bob, and Charlie, respectively, as shown in Fig.~\ref{fig3}. The
state $|\phi\rangle_{Aa} |\phi\rangle_{Bb} |\phi\rangle_{Cc}$ of
the three hybrid entanglement pairs \emph{Aa}, \emph{Bb}, and
\emph{Cc} can be rewritten as
   {\begin{eqnarray} 
|\phi_{0}\rangle&=& \frac{1}{2\sqrt{2}}\sum_{i, j, k}|\GHZ_{ijk}\rangle_{ABC}|\GHZ_{ijk}\rangle_{abc}.
 \label{Bellh3}
\end{eqnarray}}
Here the subscripts $i$, $j$, $k \in\{0, 1\}$, and
the polarization-encoded GHZ states
$|\GHZ_{ijk}\rangle_{abc}$ are defined in Eq.~(\ref{GHZnn})
for $n=3$. The eight distant stationary GHZ states among
Alice, Bob, and Charlie are of the following forms
\begin{eqnarray} 
|\GHZ_{00k}\rangle_{ABC}\!=\!\frac{1}{\sqrt{2}}\big[|\!\uparrow\rangle_{A}|\!\uparrow\rangle_{B}|\!\uparrow\rangle_{C}\!+\! (\! -1)^k\! |\!\downarrow\rangle_{A}|\!\downarrow\rangle_{B}|\!\downarrow\rangle_{C} \big], &&\nonumber\\
|\GHZ_{10k}\rangle_{ABC}\!=\!\frac{1}{\sqrt{2}}\big[|\!\downarrow\rangle_{A}|\!\uparrow\rangle_{B}|\!\uparrow\rangle_{C}\!+\! (\! -1)^k\! |\!\uparrow\rangle_{A}|\!\downarrow\rangle_{B}|\!\downarrow\rangle_{C} \big], &&\nonumber\\
|\GHZ_{01k}\rangle_{ABC}\!=\!\frac{1}{\sqrt{2}}\big[|\!\uparrow\rangle_{A}|\!\downarrow\rangle_{B}|\!\uparrow\rangle_{C}\!+\! (\! -1)^k\! |\!\downarrow\rangle_{A}|\!\uparrow\rangle_{B}|\!\downarrow\rangle_{C} \big], &&\nonumber\\
|\GHZ_{11k}\rangle_{ABC}\!=\!\frac{1}{\sqrt{2}}\big[|\!\downarrow\rangle_{A}|\!\downarrow\rangle_{B}|\!\uparrow\rangle_{C}\!+\! (\! -1)^k\!
|\!\uparrow\rangle_{A}|\!\uparrow\rangle_{B}|\!\downarrow\rangle_{C}
\big], &&\nonumber\\    
\end{eqnarray}
with $k\in\{0, 1\}$. These states constitute a complete basis for
three-QD systems.

When the ancillary node Eve performs a quantum swapping operation
with a three-photon polarization-encoded GHZ-state analysis, {the
states of the three stationary qubits, which belong to Alice, Bob,
and Charlie, are} projected into a deterministic GHZ state
according to the analysis result of Eve. That is, we can, in
principle, generate multipartite GHZ entanglement efficiently
among distant stationary qubits with a perfect efficiency.

In Sec.~\ref{sec.III}, we have described a particular pattern of
the GHZ-state analysis with a preset time delay between each two
optical elements. Now we demonstrate that the GHZ-state analysis
also works for a time-delay free pattern, by performing the
aforementioned quantum swapping as an example. Suppose both QDs in
the QND detectors are initialized in the state
$|+\rangle=(|\!\uparrow\rangle+|\!\downarrow\rangle)/\sqrt{2}$,
and all the linear-optical elements perform the same operation as
that described in Sec.~\ref{sec.III}. The three photons from
hybrid entanglement pairs \emph{Aa}, \emph{Bb}, and \emph{Cc},
subsequently pass though the analysis setup independently, rather
than transmitting them in a block pattern. After photon $a$
passing through the setup and being routed into two spatial modes that are ended with single-photon detectors, the hybrid system,
consisting of three entanglement   pairs and two QDs in the QND
detectors, evolves into
 \begin{eqnarray} 
|\phi_{1}\rangle&=&\frac{1}{2}
 \Big[|H\rangle_{a}\big(|+\rangle_{A}|+\rangle_{1}|-\rangle_{2}+|-\rangle_{A}|-\rangle_{1}|+\rangle_{2}\big)\nonumber\\
 &&-|V\rangle_{a}\big(|+\rangle_{A}|+\rangle_{1}|-\rangle_{2}-|-\rangle_{A}|-\rangle_{1}|+\rangle_{2}\big)\Big]
 \nonumber\\
 &&\otimes|\phi\rangle_{Bb}|\phi\rangle_{Cc}.
 \label{Bells3}
\end{eqnarray}
For clarity, we   assume that the standard~(destructive)
single-photon detectors work nondestructively and a photon
survives after a measurement on it, such that we can directly
specify the state of the distant stationary qubits according to
the state of the photon \emph{a}. Subsequently, the photon
\emph{b} is input into the setup when the photon \emph{a} has
passed through the setup and lead to a click of either
single-photon destructive detector  D$_1$ or D$_2$. The hybrid
system evolves into
 \begin{eqnarray} 
|\phi_{2}\rangle\!\!&=&\!\!
\frac{1}{2\sqrt{2}}\Big[\big(|\Phi^+\rangle_{AB}|+\rangle_{1}|+\rangle_{2}\nonumber\\
&&\!\!+|\Phi^-\rangle_{AB}|-\rangle_{1}|-\rangle_{2}\big)|H\rangle_{a}|H\rangle_{b}
-\big(|\Psi^+\rangle_{AB}|+\rangle_{1}|+\rangle_{2}\nonumber\\
 &&\!\! +|\Psi^-\rangle_{AB}|-\rangle_{1}|-\rangle_{2}\big)|H\rangle_{a}|V\rangle_{b}
-\big(|\Psi^+\rangle_{AB}|+\rangle_{1}|+\rangle_{2} \nonumber\\
 &&\!\! -|\Psi^-\rangle_{AB}|-\rangle_{1}|-\rangle_{2}\big)|V\rangle_{a}|H\rangle_{b}
+\big(|\Phi^+\rangle_{AB}|+\rangle_{1}|+\rangle_{2}\nonumber\\
&&\!\! -|\Phi^-\rangle_{AB}|-\rangle_{1}|-\rangle_{2}\big)|V\rangle_{a}|V\rangle_{b}
 \Big] \otimes |\phi\rangle_{Cc},
 \label{Bells41}
\end{eqnarray}
where the four Bell states of the two QDs, belonging to Alice and
Bob, 
are as follows:
\begin{eqnarray} 
|\Phi^{\pm}\rangle_{AB}=\frac{1}{\sqrt{2}}(|\!\uparrow\rangle_{A}|{\uparrow}\rangle_{B}\pm|\!\downarrow\rangle_{A}|\!\downarrow\rangle_{B}), \nonumber\\
|\Psi^{\pm}\rangle_{AB}=\frac{1}{\sqrt{2}}(|\!\uparrow\rangle_{A}|\downarrow\rangle_{B}\pm|\!\downarrow\rangle_{A}|\!\uparrow\rangle_{B}),
 \label{Bells4}
\end{eqnarray}
Now, if Eve terminates the
input of photon \emph{c} and detects the two QDs of the QND
detectors, the two distant QDs \emph{A} and \emph{B} are collapsed
to one of the Bell states given in Eq.~(\ref{Bells4}), according
to the results of the QND detectors and the measurement on photons
\emph{ab}. That is, a deterministic quantum swapping operation can
be completed between two hybrid entanglement pairs \emph{Aa} and
\emph{Bb} {by using} the passive entanglement analysis setup.

If Eve inputs the photon $c$ into the analysis setup rather than
terminating it with a measurement on the two QDs of the QND
detectors, the state $ |\phi_{2}\rangle$ of the hybrid system
evolves into the final state
\begin{eqnarray} 
|\phi_{3}\rangle&\!\!\!=\!\!\!&
\frac{1}{2\sqrt{2}}\!\sum_{ij}(\!-1)^{i+j}\!\Big[|\GHZ_{ij0}\rangle_{ABC}|\GHZ_{ij1}\rangle_{abc}|+\rangle_{1}|-\rangle_{2}\nonumber\\
&&+|\GHZ_{ij1}\rangle_{ABC}|\GHZ_{ij0}\rangle_{abc}|-\rangle_{1}|+\rangle_{2}\Big],
\;\;\;\;\;\;\;\;\;
\end{eqnarray}
with the subscripts $i,j\in\{0, 1\}$. Three distant QDs
\emph{A}, \emph{B}, and \emph{C} are projected into a
predetermined GHZ state, according to the results of the QND
detectors and the single-photon destructive detectors, when Eve
applies a polarization-encoded GHZ-state analysis on three photons
of the hybrid entangled pairs. Therefore, in principle, the
passive GHZ-state analysis works faithfully for both {cases, i.e.,
the time-delay and time-delay-free cases}, when an ideal
single-photon QND detector is available.

\section{Performance of the passive GHZ-state analyser} \label{sec.V}

\subsection{  Realistic photon scattering}

A core element of the passive GHZ-state analysis is the QND
detector for single photons. Here a unit consisting of a QD and a
micropillar cavity enables such QND detection. In principle, the
QND detector can perfectly distinguish two orthogonal polarization
states $|H\rangle$ and $|V\rangle$ of a single photon with perfect
efficiency.
However, there   are   always some imperfections that
introduce a deviation from ideal single-photon
scattering~\cite{hu2008giant, hu2008deterministic, li2018gate,
Arcari14Near-Unity, lodahl2015interfacing}, such as
 a finite single-photon bandwidth, a finite coupling $g$
between a QD and a cavity, and the nondirectional cavity side
leakage $\kappa_s$,   etc. This leads to realistic (nonideal)
scattering for a linearly polarized photon. Thus, the hybrid
system consisting of a linearly polarized single photon and a QD,
evolves as follows:
 \begin{eqnarray} 
|H\rangle|\pm\rangle&\rightarrow&\frac{1}{\sqrt{C_N}}[(r_1+r_0)|H\rangle|\pm\rangle+(r_1-r_0)|V\rangle|\mp\rangle], \nonumber\\ 
|V\rangle|\pm\rangle&\rightarrow&\frac{1}{\sqrt{C_N}}[(r_1-r_0)|H\rangle|\mp\rangle+(r_1+r_0)|V\rangle|\pm\rangle],
\;\;\;\;\;\;\;\; \label{transformMatrix17}
\end{eqnarray}
where the parameters $r$ and $r_0$ are frequency-dependent reflection coefficients
given in Eq.~(\ref{rcoeS}); $C_N=2(|r_1|^2+|r_0|^2)$ is the normalized coefficient. After scattering, the state of the
photon and the QD   evolves in two ways {independent of its
initial state}:

(1) It is flipped simultaneously with a probability
$p_1=|r_1-r_0|^2/4$, which is the desired output and it can be
simplified to perform an ideal QND detection, as given in
Eq.~(\ref{transformMatrix}), when ideal scattering with $r_1=1$
and $r_0=-1$ is achieved.

(2) The state of the photon and the QD are unchanged with the
probability $p_2=|r_1+r_0|^2/4$, which leads to errors and results
in an unfaithful QND detection for single photons.

Fortunately, this nonideal scattering does not affect the fidelity
of the passive GHZ-state analysis, since the undesired scattering
component is filtered out automatically by the PBS and only leads
to an inconclusive result rather than infidelity result by a click
of the single-photon destructive detector $D_3$.

\subsection{  Realistic analyzer efficiency}

\begin{figure}[!tpb]
\begin{center}
\includegraphics[width=8.8 cm, angle=0]{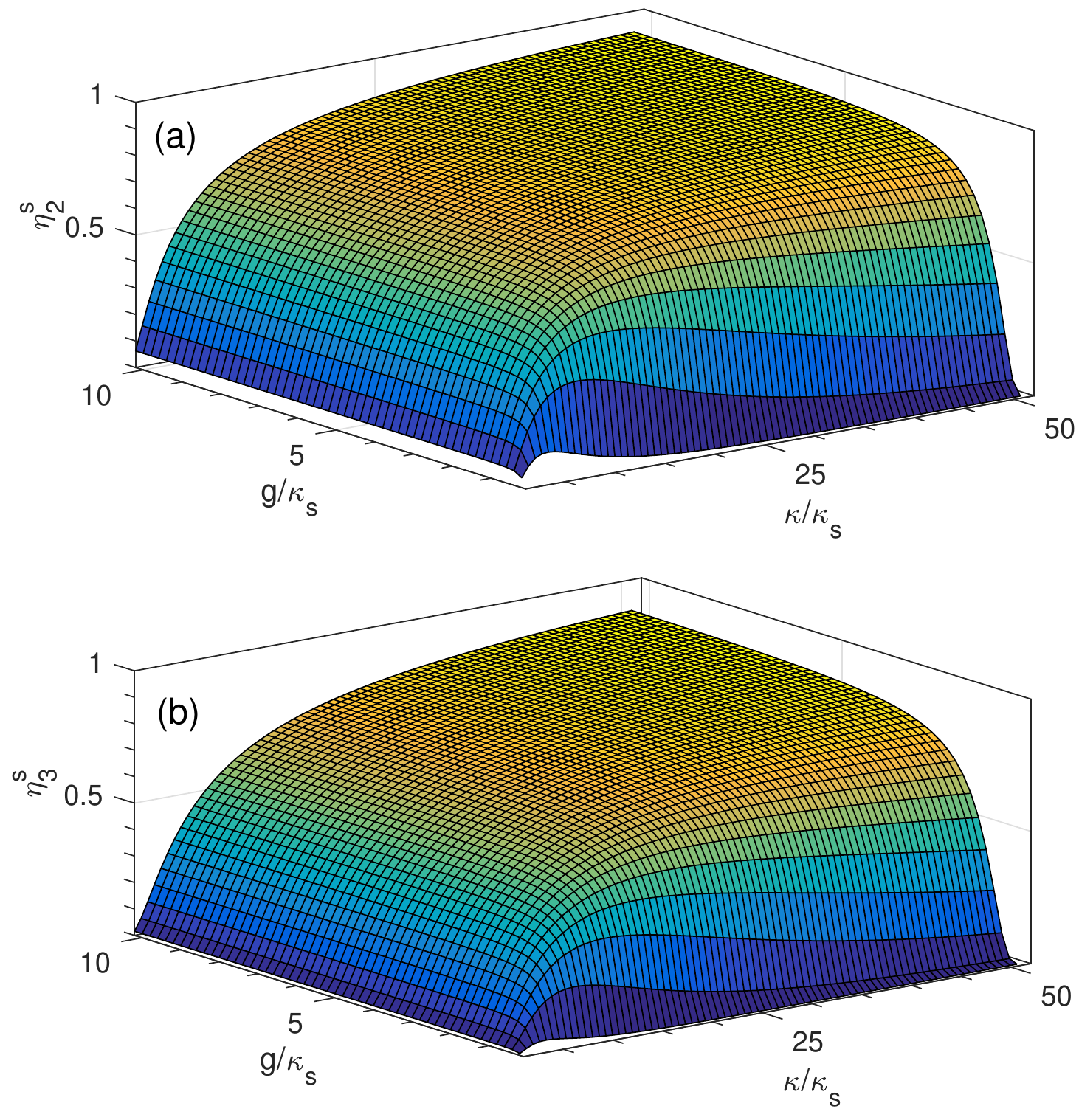}
\caption{Average efficiencies {(a) $\eta^s_2$ of the two-photon
Bell-state analyzer and (b) $\eta^s_3$ of the three-photon
GHZ-state analyzer versus the} coupling strength $g/\kappa_s$ and
the directional coupling rate of a cavity $\kappa/\kappa_s$ in
units of the cavity side-leakage rate $\kappa_s$. These averages are
calculated over all detunings of input photons, with the Gaussian
spectrum given by Eq.~(\ref{gaussianPuls}) and
$\sigma_{\omega}=\gamma$. The decay parameters are ($\kappa_s,
\gamma)=(30~\mu\eV, 0.3~\mu\eV)$. } \label{fig4}
\end{center}
\end{figure}

 {For ideal scattering, the analyzer efficiency approaches unity.
Here, we evaluate the performance of a realistic analyzer for the
general reflection amplitudes given in Eq.~(\ref{rcoeS}).}
Nonideal scattering in practical QND detection does not reduce the
fidelity of an \emph{n}-photon GHZ analysis. However,   this
realistic scattering decreases the  efficiency
$\tilde{\eta}^s_{n}$, which is defined as the probability that all photons
are detected by a single-photon destructive detector, either $D_1$
or $D_2$. For monochromatic photons of a frequency $\omega$, the
efficiency $\tilde{\eta}^s_n$ is defined as
 \begin{eqnarray} 
\tilde{\eta}^s_n={\eta^n_0\eta^n_1(\omega)}, \label{efficiency}
\end{eqnarray}
where $\eta_0$ is the efficiency of a single-photon detector $D_i$
and $\eta_1(\omega)$ is the error-free efficiency of a practical
scattering with
 \begin{eqnarray} 
\eta_1(\omega)=\left|\frac{r_1(\omega)-r_0(\omega)}{2}\right|^{2}.
\label{efficiency2}
\end{eqnarray}
The average efficiencies of the passive two-photon Bell-state and
the three-photon GHZ-state analyzers are shown in Fig.~\ref{fig4}
with decay parameters ($\kappa_s, \gamma)=(30~\mu\eV,
0.3~\mu\eV)$, which are adopted according to the QDs that are
embedded in electrically controlled cavities around
$4$~K~\cite{giesz2016coherent, somaschi2016near}. We plotted the
average efficiencies $\eta^s_2$ and $\eta_3^s$ versus the coupling
strength $g/\kappa_s$ and the directional coupling rate of the
cavity $\kappa/\kappa_s$ for a given Gaussian single-photon pulse
defined by the spectrum
\begin{eqnarray}
f(\omega)=\frac{1}{\sqrt{\pi}\sigma_{\omega}}\exp\left[-\left(\frac{\omega-\omega_c}{\sigma_{\omega}}\right)^2\right],
\label{gaussianPuls}
\end{eqnarray}
where $\omega_c$ is the center frequency and $\sigma_{\omega}$
denotes the pulse bandwidth with $\omega_c=\omega_{X^-}$ and
$\sigma_{\omega}=\gamma$.  {Here the average efficiencies are
calculated in the frequency domain. The reflection coefficients
appear as a frequency-dependent redistribution function that is
proportional to $|r_1(\omega)-r_0(\omega)|^{2n}$ as
follows~\cite{DiVincenzo_2013,Cohen2018}:
\begin{eqnarray}
\eta^s_n=\int{}d\omega{}f(\omega)\eta^n_0\left|\frac{r_1(\omega)-r_0(\omega)}{2}\right|^{2n}.
\label{gaussianPulsEta}
\end{eqnarray}}

\begin{figure}[!tpb]
\begin{center}
\includegraphics[width=8.8 cm, angle=0]{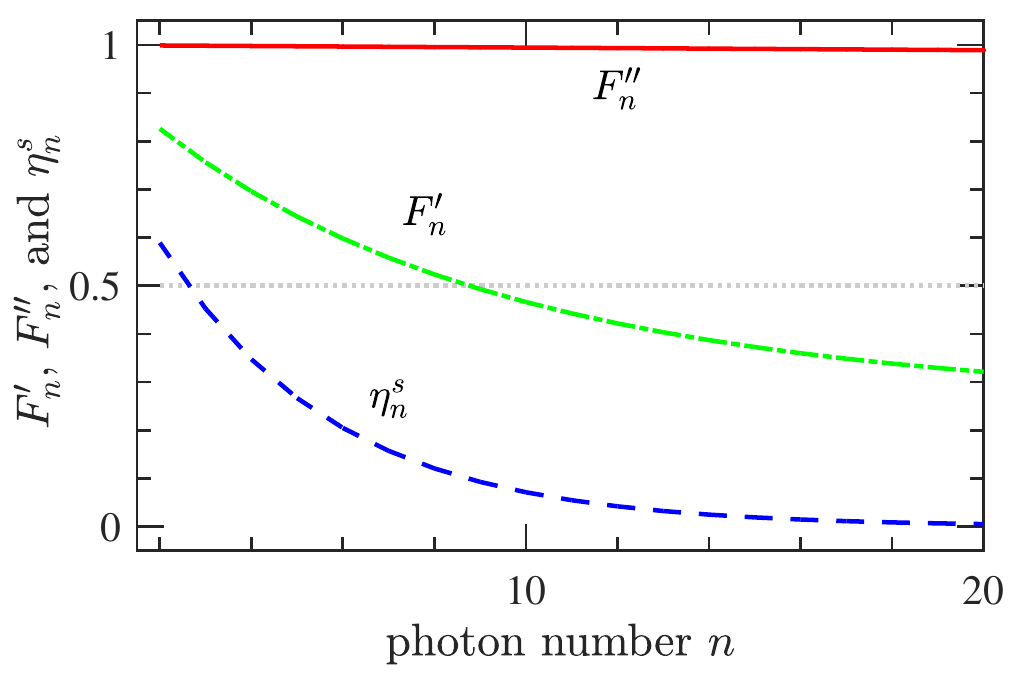}
\caption{ {Average fidelities $F'_n$, $F''_n$, and efficiency $\eta_n^s$ versus photon number $n$.  Here $F'_n\equiv F_n(T_2=10.9~{\rm ns})$ and $F''_n\equiv F_n(T_2=2~{\rm \mu{}s})$. These averages are calculated over all detunings of input photons, with the Gaussian
spectrum given by Eq.~(\ref{gaussianPuls}) and
$\sigma_{\omega}=2\gamma$. The decay parameters are ($\kappa_s,\gamma)=(30~\mu\eV, 0.3~\mu\eV)$, $g=\kappa_s$, and $\kappa=9\kappa_s$ with $C=10$. Meanwhile, the line of the fidelities at $1/2$ is shown for reference.}} \label{fig5}
\end{center}
\end{figure}

In general, the average efficiencies of the passive two-photon
Bell-state and three-photon GHZ-state analyzers increase when the
coupling $g/\kappa_s$ between a QD and a cavity is increased for a
given directional coupling rate $\kappa/\kappa_s$. This is because
the cooperativity
\begin{eqnarray}
C=\frac{g^2}{\gamma\kappa_T}=\frac{g^2}{\gamma(\kappa+\kappa_s)},
\end{eqnarray}
which is defined as an essential
parameter quantifying the loss of an atom-cavity system, increases
when we increase $g/\kappa_s$ and keep other parameters unchanged.
 {For a given $g/\kappa_s$, the average efficiencies of
these two analyzers first increase and then decrease when
$\kappa/\kappa_s$ is increased, as shown in Fig.~\ref{fig4}. This is mainly
due to the competition between an increased ratio of
$\kappa/\kappa_s$ and a decreased cooperativity $C$.}
Therefore, one can maximize the efficiencies by using
cavities  with a mediate $\kappa$, which can be achieved,
e.g., by decreasing the number of the Bragg reflector of a
micropillar cavity. For simplicity, we set the efficiency of a
single-photon destructive detector as $\eta_0=1$.

For the two-photon Bell-state analyzer, its average efficiency
$\eta^s_2=0.304$ for an experimental demonstrated coupling
$g/\kappa_s=1$ and the directional coupling rate of a cavity,
$\kappa/\kappa_s=3$, which corresponds to a cooperativity $C=25$.
For the three-photon GHZ-state analyzer, one can obtain the
average efficiency $\eta^s_3\simeq0.168$ for the same systematic
parameters. If $\kappa$ is increased to $\kappa/\kappa_s=19$ with
a cooperativity $C=5$~\cite{giesz2016coherent},
 the average efficiencies are increased to $\eta^s_2\simeq0.664$ and
$\eta^s_3\simeq0.541$, respectively.
 {Note that the adiabatic condition  is still satisfied in this case, since the photon bandwidth $\sigma_{\omega}=\gamma$ is much smaller than $2g^2/\kappa_T=2C\gamma$.}
The protocol works with a higher efficiency for analyzing photons with a narrower
bandwidth. However, this, in turn, usually increases the time
period of the scattering process, and, thus, decreases the
analyzer fidelity limited by  QD decoherence~\cite{hu2008deterministic,OBrien2016}. The fidelity of our analyzer is given by
\begin{eqnarray}
F_n(T_2)=\frac{1}{4}\big[1+\exp(-t_n/T_2)\big]^2, \label{Fn}
\end{eqnarray}
which is determined by the process for distinguishing $|\GHZ_{i_1i_2...0}\rangle$ from $|\GHZ_{i'_1i'_2...1}\rangle$
(measuring $X$-type stabilizer $S_1$), since the process for measuring
$Z$-type stabilizers is completed directly  by single-photon detectors  and is independent of QD decoherence.
 {Here, the time required to complete the $n$-photon GHZ-state analysis is given by $t_n \simeq n t_0$. In our numerical calculations shown in Fig.~\ref{fig5} and Table~\ref{nphon}, we assumed $t_0\simeq1.10$ ns for performing a single-photon scattering
process with a bandwidth $\sigma_{\omega}=2\gamma$. Moreover, $T_2$ is the coherence time of the electron spin in a QD.}

 {Typically, the coherence time $T_2$ is in the $1$--$10$
ns range~\cite{warburton2013single,de2013ultrafast,mikkelsen2007optically}.  Taking an experimental
accessible value of $T_2=10.9$~ns~\cite{mikkelsen2007optically}, we obtain the
corresponding analyzer fidelity $F'_n\equiv F_n(T_2=10.9~{\rm
ns})$ versus the photon number $n$ for the parameters ($g,
\kappa_s, \gamma)=(30~\mu\eV, 30~\mu\eV, 0.3~\mu\eV)$ and
$\kappa/\kappa_s=9$ with $C=10$, as shown by the green dash-dot
curve in Fig.~\ref{fig5} and listed Table~\ref{nphon}. For a four-photon system, the
analyzer fidelity can reach $F'_4\simeq0.70$.  For an eight-photon
system, $F'_{8}$ is still larger than $0.5$.}

 {The fidelity of our GHZ-state analyzer is influenced by the coherence time $T_2$.
Note that the coherence time $T_2$ can be optimized and improved
to be longer than $2\;\mu$s, when the high-degree nuclear-spin
bath polarization or spin-echo refocusing methods are applied~~\cite{warburton2013single,de2013ultrafast}. When $T_2=2$ $\mu$s, we can achieve the fidelity
$F''_n\equiv F_n(T_2=2~{\rm \mu{}s})>0.988$ for $n\leq 20$~{(see
the red solid curve in Fig.~\ref{fig5})} assuming all the other parameters to
be the same as for $F'_n$ in this figure.}

 {In contrast to the fidelity, the average efficiency
$\eta_n^s$ is independent of $T_2$, because  the effective output
component, which is involved in a scattering process, is
independent of the state of the QD, as shown in Eq.~(\ref{transformMatrix17}). In
general, $\eta_n^s$ decreases when the photon number $n$ increases
(see the blue dashed curve in Fig.~\ref{fig5} and Table~\ref{nphon}). For a
20-photon system, the average efficiency of our protocol is
equal to $\eta^s_{20}=0.0039$, which is many orders higher than
the efficiency given by $1/2^{(n-1)}=2^{-19}$ for the standard
analyzers consisting of linear-optical elements and single-photon
detectors~\cite{Pan1998GHZanalyzer}.}

\begin{table}
 \centering
\begin{tabular}{c c c c | c c c c}
 \hline
 \hline
 ~$n$~ & ~~$F'_n$~~~& ~~~$F''_n$~~~ & ~~~$\eta_n^s$~~ & ~$n$~ & ~~~$F'_n$~~~& ~~~$F''_n$~~~ & ~~~$\eta_n^s$~~~\\
 \hline
 2& 0.826 & 0.9989  & 0.5893  &  6&0.5981 & 0.9967  &0.2047 \\
 3&0.7564  & 0.9984 &0.4524  &7 &0.5583  &0.9962 &0.1572  \\
 4&0.6961 & 0.9978  &0.3473 &8 &0.5235 &0.9956 &0.1207 \\
 5&0.6437  & 0.9973 &0.2666  &20 &0.3141 &0.9886 &0.0039\\
 \hline
 \hline
\end{tabular}
\caption{ {Average fidelities $F'_n$, $F''_n$, and efficiency $\eta_n^s$
versus photon number $n$ for the parameters assumed in Fig.~\ref{fig5}. Here $F'_n\equiv F_n(T_2=10.9~{\rm ns})$ and $F''_n\equiv F_n(T_2=2~{\rm \mu{}s})$.}}\label{nphon}
\end{table}

\section{Discussion} \label{sec.VI}

The linear-optical implementation of the GHZ-state analyzer
 passively distinguishes two GHZ states
$|\GHZ_{00...0}\rangle$ and $|\GHZ_{00...1}\rangle$ 
from the remaining $(2^n-2)$ GHZ states, and
enables a complete analysis for $2^n$ GHZ states when  many
ancillary photons and detectors are used~\cite{Kop2007linear}.
This kind of   GHZ analysis is much like a GHZ-state generation
that is constructed by linear optics and
post-selection~\cite{pan2012multiphoton, Krenn2017PathGHZ,
Bergamasco2017pathGHZ}. Currently, the {GHZ state of a 10-photon
system has been demonstrated by using linear
optics}~\cite{wang10photonGHZ,chen2017observation}. The existing
GHZ-state analyzers, which are based on optical nonlinearities,
have been proposed by cascading two-photon parity QND
detectors~\cite{Sheng2008EPP, Pittman2002PCDCNOT}. Such an
analyzer can, in principle, distinguish $2^n$ GHZ states of an
\emph{n}-photon system nondestructively, when it is assisted by
fast switching and/or active operations during the entangled-state
analysis. These operations dramatically increase its experimental
complexity and consume more quantum resources. Furthermore, such
implementations always require a strong optical nonlinearity to
keep the analysis faithful.

Our scheme of a passive GHZ-state analysis for \emph{n}
polarization-encoded photons uses only linear-optical elements,
and single-photon destructive and nondestructive detectors. This
analyzer can, in principle, deterministically distinguish among
$2^n$ GHZ states for \emph{n}-photon systems,  and hence it
combines the advantages of those based on linear optics with those
based on optical nonlinearities. Moreover, our scheme eliminates
disadvantages of such standard analyzers by designing an
error-tolerant QND detection for single photons, and can be useful for efficient
implementations of all-photonic quantum repeaters, even including
those without quantum memory~\cite{Hasegawa2019,Li2019}.

The proposed QND detector   consists of a four-level emitter
coupled to a microcavity or waveguide~\cite{reithmaier2004strong,
hu2008giant,li2018gate, hu2008deterministic, Arcari14Near-Unity,
lodahl2015interfacing}, such as a negatively charged QD coupled to
a micropillar cavity.   This analyzer is also compatible with the
proposals of realistic QND
detection~\cite{witthaut2010photon, hu2008giant,OBrien2016,
reiserer2013nondestructive} for single photons; {however, as we
have shown, it is more efficient than {the standard ones for
several reasons: Our QND detector can work in a passive way} and
can faithfully distinguish photon numbers subsequently passing
through it in an  {even-odd} basis. Furthermore, it is
error-tolerant, when it is used to detect linearly polarized
photons.

Our description of scattering imperfections includes the following: finite
photon-pulse bandwidth $\sigma_{\omega}$, cavity loss $\kappa_s$,
and finite coupling $g$. {This realistic scattering process leads
to a} hybrid entangled state of a photon and a QD, consisting of
ideal scattering component and the error scattering component.
{When the two QDs couple equally to their respective micropillar
cavities, the} error component is passively filtered out by a PBS
and then is heralded by a click of a single-photon destructive
detector, leading to an inconclusive result rather than an
unfaithful GHZ-state analysis. {In practice, the two QDs might be
different due to inhomogeneous broadening and could couple
differently to their respective cavities~\cite{warburton2013single}. This would lead to
different scattering processes, which result in different hybrid
entangled states of a photon and a QD, when a photon is reflected
by different QND detectors. This effect, in principle, can be
suppressed by inserting a passive modulator before the QND
detector with a larger ideal scattering component and by tuning it
to match that of the other QND detector~\cite{li2014distillation}.
}

{Furthermore, a QD is a candidate for quantum information
processing due to its very good characteristics concerning its
optical initialization, single-qubit manipulation, and readout,
based on well-developed semiconductor
technologies~\cite{liu2010quantum, buluta2011natural,
de2013ultrafast, lodahl2017quantum,DingPhoton2016PRL}.}
{The coherence time of a QD electron spin can be several microseconds
at temperatures around 4~K~\cite{warburton2013single,de2013ultrafast}, while a
single-photon scattering is  accomplished on  nanosecond
timescales. Moreover, the present protocol of entanglement
analysis can be generalized to other systems with a required level
structure~\cite{schwartz2016deterministic,buterakos2017deterministic}.}
Recently, a five-photon polarization-encoded cluster state has
been demonstrated with a confined dark exciton in a
QD~\cite{schwartz2016deterministic} and an all-photonic quantum
repeater protocol was described with a similar solid-state
four-level emitter~\cite{buterakos2017deterministic}.

\section{conclusions} \label{sec.VII}
In summary, we proposed a resource-efficient analyzer of {Bell} and
Greenberger-Horne-Zeilinger states of multiphoton systems.
Quantum-nondemolition detection is implemented in our analyzer
with two four-level emitters (e.g., quantum dots), each coupled to
a one-dimensional system (such as optical micropillar cavity or a
photonic nanocrystal waveguide). This QND  measures the number of photons
passing through it in the even-odd basis and constitutes a faithful element for the
GHZ-state analyzer by introducing a passively
error-filtering structure with linear optics.

The main idea of our proposal can be simply explained in
terms of stabilizers for GHZ states defined in
Eq.~(\ref{stabilizers}). Specifically, we proposed to measure the
parity of the $X$-type stabilizer [$k=1$ in
Eq.~(\ref{stabilizers})] with two quantum dots and to measure the
parity of the $Z$-type stabilizers [$k=2,3,...,n$ in
Eq.~(\ref{stabilizers})] with direct polarization measurements on
each photon scattered by the QDs.
There are  neither active operations  nor adaptive switching in the proposed method,
since the faithful GHZ-state analysis for multiple photon systems
works efficiently by passively arranging two QND detectors,
single-photon destructive detectors, and linear-optical elements.
Furthermore, the {described method is universal, as it enables
two-photon Bell-state and multiphoton GHZ-state analyses.} All
these distinct characteristics make the proposed passive analyzers
simple and resource efficient for long-distance multinode quantum
communication and quantum networks.

\section*{Acknowledgments}

This work was supported in part by the National Key R\&D Program
of China (Grant No. 2017YFA0303703), the Natural Science
Foundation of Jiangsu Province (Grant No. BK20180461),  the
National Natural and Science Foundation of China (Grants Nos.
No. 11874212,  No. 11574145, No.
11890700,  No. 11890704,  No. 11904171, and No. 11690031),
and the Fundamental Research Funds for the Central Universities (Grant No. 021314380095). 
A.M. and F.N.
acknowledge a grant from the John Templeton Foundation. F.N. is
also supported in part by the: MURI Center for Dynamic
Magneto-Optics via the Air Force Office of Scientific Research
(AFOSR) (Grant No. FA9550-14-1-0040), Army Research Office (ARO) (Grant No.
W911NF-18-1-0358), Asian Office of Aerospace Research
and Development (AOARD) (Grant No. FA2386-18-1-4045), Japan
Science and Technology Agency (JST) (Q-LEAP program  and CREST
Grant No. JPMJCR1676), Japan Society for the Promotion of Science
(JSPS) (JSPS-RFBR Grant No. 17-52-50023, and JSPS-FWO Grant No.
VS.059.18N),  the NYY PHI Labs, and the RIKEN-AIST Challenge Research Fund.

\appendix
\section{Analyzer of two-photon polarization-encoded Bell states}\label{app-Bell}
\renewcommand{\theequation}{\thesection\arabic{equation}}

 {Here we give a pedagogical example of our method limited to
the polarization-encoded Bell-state analysis.}

 {The passive analyzer, in principle, enables a deterministic analysis of two-photon polarization-encoded Bell states.
For any two-photon system, the four Bell states can be described as follows,
\begin{eqnarray} 
 |\phi^{\pm}\rangle&=&\frac{1}{\sqrt{2}}\left(|H\rangle|H\rangle\pm|V\rangle|V\rangle\right), \nonumber\\
 |\psi^{\pm}\rangle&=&\frac{1}{\sqrt{2}}\left(|H\rangle|V\rangle\pm|V\rangle|H\rangle\right).
\label{Bell-app}
\end{eqnarray}
Photon pairs in these states, after passing through the analyzer,
lead to four different results that are heralded by single-photon destructive and QND detectors.}

\begin{table}[!h]
 \centering
\begin{tabular}{ c c c c c }
 \hline
 \hline
 & $\ket{HH}/\ket{VV}$ & $\ket{HV}/\ket{VH}$ &~~$\ket{++}$~~&~~$\ket{--}$~~\\
 \hline
$|\phi^{+}\rangle$ & $\surd$ & & $\surd$ & \\
$|\phi^{-}\rangle$ & $\surd$ & & & $\surd$ \\
$|\psi^{+}\rangle$ & & $\surd$ & $\surd$ &\\
$|\psi^{-}\rangle$ & & $\surd$ & &$\surd$ \\
 \hline
 \hline
\end{tabular}
\caption{ Complete two-photon Bell-state analysis. Here $\ket{ij}$
represents the measurement result of the two single-photon
detectors (QDs) with $|ij\rangle=\{|HH\rangle, |VV\rangle,
|HV\rangle, |VH\rangle\}$ ($|ij\rangle=\{|++\rangle,
|--\rangle\}$). We use the standard notation for the Bell states
$\ket{\phi^{\pm}}$ and $\ket{\psi^{\pm}}$, as given in
Eq.~(\ref{Bell-app}).}\label{n2}
\end{table}

 {Suppose now that the QD in each QND detector is initialized to the
state $|+\rangle$. The HWP introduces a  Hadamard transformation
on photons passing it, i.e.,
$|H\rangle\rightarrow(|\!H\rangle+|\!V\rangle)/\sqrt{2}$, or
$|V\rangle\rightarrow(|\!H\rangle-|\!V\rangle)/\sqrt{2}$, and
evolves the states $|\phi^{+}\rangle$, $|\phi^{-}\rangle$,
$|\psi^{+}\rangle$, and $|\psi^{-}\rangle$ into
$|\psi_1\rangle=|\phi^{+}\rangle$,
$|\psi_2\rangle=|\psi^{+}\rangle$,
$|\psi_3\rangle=|\phi^{-}\rangle$, and
$|\psi_4\rangle=-|\psi^{-}\rangle$, respectively. The original
states $|\phi^{+}\rangle$ and $|\psi^{+}\rangle$, with a relative
phase of \emph{zero}, are changed into states consisting of even
numbers of $V$-polarized photons, i.e., $|H\rangle|H\rangle$ and
$|V\rangle|V\rangle$. While the original states $|\phi^{-}\rangle$
and $|\psi^{-}\rangle$ with a relative phase of \emph{$\pi$} are
changed into states consisting of odd numbers of $V$-polarized
photons, i.e., $|H\rangle|V\rangle$ and $|V\rangle|H\rangle$.}

 {Subsequently, photons in the $V$-polarized ($H$-polarized)
states are reflected (transmitted) by the PBS, and are scattered
by the detector QND$_1$ (QND$_2$). Photon pairs in the states
$|\psi_1\rangle$, $|\psi_2\rangle$, $|\psi_3\rangle$, and
$|\psi_4\rangle$, which are combined with two QDs, are changed
into the states
\begin{eqnarray}
|\psi'_1\rangle&=&|\phi^{+}\rangle|+\rangle_1|+\rangle_2,  \nonumber\\
|\psi'_2\rangle&=&|\psi^{+}\rangle|-\rangle_1|-\rangle_2,  \nonumber\\
|\psi'_3\rangle&=&-|\phi^{-}\rangle|+\rangle_1|+\rangle_2,  \nonumber\\
|\psi'_4\rangle&=&|\psi^{-}\rangle|-\rangle_1|-\rangle_2.
\end{eqnarray}
The original Bell states with  relative phases \emph{zero} and
$\pi$ can be distinguished from each other, according to the
states of the QDs.}

 {To read out the original polarization information of the photon pair,
the HWP between two PBSs introduces a  Hadamard transformation on
photons passing through it and transforms  $|\psi'_i\rangle$ into
$|\psi''_i\rangle$ for $i=1, 2, 3, 4$, with
\begin{eqnarray}
|\psi''_1\rangle&=&|\phi^{+}\rangle|+\rangle_1|+\rangle_2, \nonumber\\
|\psi''_2\rangle&=&|\phi^{-}\rangle|-\rangle_1|-\rangle_2, \nonumber\\
|\psi''_3\rangle&=&-|\psi^{+}\rangle|+\rangle_1|+\rangle_2,  \nonumber\\
|\psi''_4\rangle&=&-|\psi^{-}\rangle|-\rangle_1|-\rangle_2,
\end{eqnarray}
which transforms the photon pair state to their original state, up
to an overall phase $\pi$. Therefore, we can distinguish
$|\psi''_1\rangle$ and $|\psi''_2\rangle$ from $|\psi''_3\rangle$
and $|\psi''_4\rangle$ by performing single-photon destructive
measurements in  the vertical-horizontal basis. Thus, one can
distinguish $ |\phi^{\pm}\rangle$ from $ |\psi^{\pm}\rangle$.
Finally, we can completely  identify the four Bell states by the
measurement results of the single-photon destructive  and QND
detectors, as shown in Table~\ref{n2}.}

\section{Analyzer of three-photon polarization-encoded GHZ states}\label{app-ghz3}}
\renewcommand{\theequation}{B\arabic{equation}}

\begin{table}
 \centering
\begin{tabular}{c c c c c c c }
 \hline
 \hline
 & ~~C1~~ & ~~C2~~ & ~~C3~~ & ~~C4~~ & ~$\ket{+-}$~ & ~$\ket{-+}$~ \\
 \hline
$|\GHZ_{000}\rangle$ & $\surd$ & & & & $\surd$ & \\
$|\GHZ_{001}\rangle$ & $\surd$ & & & & & $\surd$ \\
$|\GHZ_{100}\rangle$ & & $\surd$ & & & $\surd$ & \\
$|\GHZ_{101}\rangle$ & &$\surd$ & & & & $\surd$ \\
$|\GHZ_{010}\rangle$ & & & $\surd$ & & $\surd$ & \\
$|\GHZ_{011}\rangle$ & & & $\surd$ & & & $\surd$ \\
$|\GHZ_{110}\rangle$ & & & & $\surd$ & $\surd$ & \\
 $|\GHZ_{111}\rangle$ & & & &$\surd$ & & $\surd$ \\
 \hline
 \hline
\end{tabular}
\caption{ Complete three-photon GHZ-state analysis. Here the
measurement results C1, C2, C3, and C4 of the two single-photon
detectors D$_{1}$ and D$_{2}$ correspond to either $\ket{HHH}$ or $\ket{VVV}$,
 and similarly for $\ket{HVV}$ or $\ket{VHH}$, $\ket{VHV}$ or $\ket{HVH}$, and
$\ket{VVH}$ or $\ket{HHV}$, respectively. Here, $|+-\rangle$ and $|-+\rangle$
denote two possible results of the measurement on the two QDs.
}\label{n3}
\end{table}

 {Here we give another pedagogical example of our method of
state analysis for polarization-encoded three-photon GHZ states.}

 {For a three-photon system, the eight GHZ states can be
written as follows,
\begin{eqnarray} 
|\GHZ_{00k}\rangle_{abc}\!=\!\frac{1}{\sqrt{2}}\big[|\!{H}\rangle_{a}|\!{H}\rangle_{b}|\!{H}\rangle_{c}\!+\! (\! -1)^k\! |\!{V}\rangle_{a}|\!{V}\rangle_{b}|\!{V}\rangle_{c} \big], &&\nonumber\\
|\GHZ_{10k}\rangle_{abc}\!=\!\frac{1}{\sqrt{2}}\big[|\!{V}\rangle_{a}|\!{H}\rangle_{b}|\!{H}\rangle_{c}\!+\! (\! -1)^k\! |\!{H}\rangle_{a}|\!{V}\rangle_{b}|\!{V}\rangle_{c} \big], &&\nonumber\\
|\GHZ_{01k}\rangle_{abc}\!=\!\frac{1}{\sqrt{2}}\big[|\!{H}\rangle_{a}|\!{V}\rangle_{b}|\!{H}\rangle_{c}\!+\! (\! -1)^k\! |\!{V}\rangle_{a}|\!{H}\rangle_{b}|\!{V}\rangle_{c} \big], &&\nonumber\\
|\GHZ_{11k}\rangle_{abc}\!=\!\frac{1}{\sqrt{2}}\big[|\!{V}\rangle_{a}|\!{V}\rangle_{b}|\!{H}\rangle_{c}\!+\! (\! -1)^k\! |\!{H}\rangle_{a}|\!{H}\rangle_{b}|\!{V}\rangle_{c}
\big]. && \nonumber\\    
\end{eqnarray}
To distinguish these eight GHZ states from one another, we input
photons \emph{(a,b,c)} into the setup for the GHZ-state analysis.
Photons \emph{(a,b,c)} in the GHZ states
$|\GHZ_{ijk}\rangle_{abc}$, $i, j, k\in\{0, 1\}$ pass through the
HWP that performs a Hadamard operation on them, and are changed,
respectively, into the states
\begin{eqnarray} 
|\Phi_{ij0}\rangle_{abc}\!=\!\frac{1}{2}\big(|G^{ij}_0\rangle+\sqrt{3}|G^{ij}_2\rangle), &&\nonumber\\
|\Phi_{ij1}\rangle_{abc}\!=\!\frac{1}{2}\big(\sqrt{3}|G^{ij}_1\rangle+|G^{ij}_3\rangle),
&& 
\end{eqnarray}
where the ancillary states $|G^{ij}_m\rangle$ with $m=0, 1, 2, 3,$
are given in Sec.~\ref{GHZ-state-analysis} and can be detailed as
follows:
\begin{eqnarray} 
|G^{ij}_0\rangle\!&=&\!|\!{H}\rangle_{a}|\!{H}\rangle_{b}|\!{H}\rangle_{c}, \nonumber\\
|G^{ij}_1\rangle\!&=&\! \frac{\sigma^{i}_{z_{a}}\otimes\sigma^{j}_{z_{b}}}{\sqrt{3}}\big( |\!{V}\rangle_{a}|\!{H}\rangle_{b}|\!{H}\rangle_{c}\nonumber\\
&&+ |\!{H}\rangle_{a}|\!{V}\rangle_{b}|\!{H}\rangle_{c}+ |\!{H}\rangle_{a}|\!{H}\rangle_{b}|\!{V}\rangle_{c}\big),\nonumber\\
|G^{ij}_2\rangle\!&=&\! \frac{\sigma^{i}_{z_{a}}\otimes\sigma^{j}_{z_{b}}}{\sqrt{3}}\big( |\!{H}\rangle_{a}|\!{V}\rangle_{b}|\!{V}\rangle_{c}\nonumber\\
&&+ |\!{V}\rangle_{a}|\!{H}\rangle_{b}|\!{V}\rangle_{c}+ |\!{V}\rangle_{a}|\!{V}\rangle_{b}|\!{H}\rangle_{c}\big),\nonumber\\
|G^{ij}_3\rangle\!&=&\!\sigma^{i}_{z_{a}}\otimes\sigma^{j}_{z_{b}}|\!{V}\rangle_{a}|\!{V}\rangle_{b}|\!{V}\rangle_{c}. 
\end{eqnarray}}

 {The GHZ states $|\GHZ_{ij0}\rangle_{abc}$
($|\GHZ_{ij1}\rangle_{abc}$) with the relative phase $0$~($\pi$) can
be distinguished from one another by measuring the numbers of
$V$-polarized photons in the even-odd basis with QND detectors.
The QND detectors initialized to the state $|+\rangle$  flip the
states of the QD and photon during the scattering process, and
evolve photons \emph{(a,b,c)} and two QDs into the states:
\begin{eqnarray} 
|\Phi'_{ij0}\rangle_{abc}\!=\!\frac{\sigma_{x_{a}}\otimes\sigma_{x_{b}}\otimes\sigma_{x_{c}}}{2}\big(|G^{ij}_0\rangle+\sqrt{3}|G^{ij}_2\rangle\big)
|+\rangle_1|-\rangle_2, &&\nonumber\\
|\Phi'_{ij1}\rangle_{abc}\!=\!\frac{\sigma_{x_{a}}\otimes\sigma_{x_{b}}\otimes\sigma_{x_{c}}}{2}\big(\sqrt{3}|G^{ij}_1\rangle+|G^{ij}_3\rangle\big)
|-\rangle_1|+\rangle_2.\nonumber\\
\end{eqnarray}
The original GHZ states with  relative phases $0$ and
$\pi$ can be distinguished from one another, according to the
states of the QDs.}

 {To read out the original polarization information of  photons
\emph{(a,b,c)}, the HWP between two PBSs introduces a  Hadamard
transformation on photons passing through it and transforms
$|\Phi'_{ijk}\rangle$ into $|\Phi''_{ijk}\rangle$ with
\begin{eqnarray}
|\Phi''_{ij0}\rangle_{abc}\!=\! (-1)^{i+j}|\GHZ_{ij1}\rangle_{abc}|+\rangle_1|-\rangle_2, &&\nonumber\\
|\Phi''_{ij1}\rangle_{abc}\!=\! (-1)^{i+j}|\GHZ_{ij0}\rangle_{abc}|-\rangle_1|+\rangle_2.
\end{eqnarray}
Now the photons \emph{(a,b,c)}  are transformed to their original
state, up to a phase difference $\pi$, which is independent of the
results of the single-photon destructive measurements in  the
vertical-horizontal basis. Therefore, we can  completely  identify
the eight GHZ states by the measurement results of the
single-photon destructive  and QND detectors, as shown in
Table~\ref{n3}.}


%

\end{document}